\documentclass[runningheads]{llncs}
\usepackage{amssymb}
\usepackage{graphicx}
\usepackage{llncsdoc}
\usepackage{makeidx, epsfig,amsmath}  
\usepackage{url}

\usepackage[colorinlistoftodos]{todonotes}
\usepackage[colorlinks=true, allcolors=blue]{hyperref}
\usepackage[title]{appendix}
\usepackage{amsfonts}
\usepackage{multirow}
\usepackage{booktabs}

\usepackage{subcaption}
\usepackage{epsfig}
\usepackage{graphicx}
\usepackage{amsmath}
\usepackage{amssymb}
\usepackage{epsfig}
\usepackage{epstopdf}
\usepackage{wrapfig}
\usepackage{multirow}
\usepackage{array}
\usepackage{tabulary}
\usepackage{graphicx}
\usepackage{color}
\usepackage{makecell}
\usepackage{marvosym}
\renewcommand{\thefootnote}{*}
\makeatletter\def\Hy@Warning#1{}\makeatother

\newcolumntype{K}[1]{>{\centering\arraybackslash}p{#1}}

\begin{document}
%
%

\title{Cluster-Induced Mask Transformers for Effective Opportunistic Gastric Cancer Screening on Non-contrast CT Scans}

\author{Mingze Yuan$^{1,2,3,*}$, Yingda Xia$^{1,}$\textsuperscript{\Letter}, Xin Chen$^{4,}$\textsuperscript{\Letter}, Jiawen Yao$^{1,3}$, Junli Wang$^{5}$, Mingyan Qiu$^{1,3}$, Hexin Dong$^{1,2,3}$, Jingren Zhou$^{1}$, Bin Dong$^{2,6}$, Le Lu$^{1}$, Li Zhang$^{2}$, Zaiyi Liu$^{4,}$\textsuperscript{\Letter}, \and Ling Zhang$^{1}$}
\authorrunning{Yuan, M. et al.}
\titlerunning{Effective Opportunistic Gastric Cancer Screening}

\institute{$^{1}$DAMO Academy, Alibaba Group
$^{2}$Peking University\\
$^{3}$Hupan Lab, 310023, Hangzhou, China \\ 
$^{4}$Guangdong Province People's Hospital \\
$^{5}$The First Affiliated Hospital of Zhejiang University \\
$^{6}$Peking University Changsha Institute for Computing and Digital Economy
}

\maketitle              

\begin{abstract}
Gastric cancer is the third leading cause of cancer-related mortality worldwide, but no guideline-recommended screening test exists. Existing methods can be invasive, expensive, and lack sensitivity to identify early-stage gastric cancer. In this study, we explore the feasibility of using a deep learning approach on non-contrast CT scans for gastric cancer detection. We propose a novel cluster-induced Mask Transformer that jointly segments the tumor and classifies abnormality in a multi-task manner. Our model incorporates learnable clusters that encode the texture and shape prototypes of gastric cancer, utilizing self- and cross-attention to interact with convolutional features. In our experiments, the proposed method achieves a sensitivity of 85.0\% and specificity of 92.6\% for detecting gastric tumors on a hold-out test set consisting of 100 patients with cancer and 148 normal. In comparison, two radiologists have an average sensitivity of 73.5\% and specificity of 84.3\%. We also obtain a specificity of 97.7\% on an external test set with 903 normal cases. Our approach performs comparably to established state-of-the-art gastric cancer screening tools like blood testing and endoscopy, while also being more sensitive in detecting early-stage cancer. This demonstrates the potential of our approach as a novel, non-invasive, low-cost, and accurate method for opportunistic gastric cancer screening.
\footnotetext{Work was done during an internship at DAMO Academy, Alibaba Group.}
\renewcommand{\thefootnote}{\Letter}
\footnotetext{Corresponding authors: yingda.xia@alibaba-inc.com; \{wolfchenxin, zyliu\}@163.com}
\end{abstract}
\keywords{Gastric cancer $\cdot$ Large-scale cancer screening $\cdot$ Mask Transformers $\cdot$ Non-contrast CT}

\section{Introduction}
Gastric cancer (GC) is the third leading cause of cancer-related deaths worldwide~\cite{smyth2020gastric}. The five-year survival rate for GC is approximately 33\% \cite{seer2022cancer}, which is mainly attributed to patients being diagnosed with advanced-stage disease harboring unresectable tumors. This is often due to the latent and nonspecific signs and symptoms of early-stage GC. However, patients with early-stage disease have a substantially higher five-year survival rate of around 72\% \cite{seer2022cancer}. Therefore, early detection of resectable/curable gastric cancers, preferably before the onset of symptoms, presents a promising strategy to reduce associated mortality. Unfortunately, current guidelines do not recommend any screening tests for GC \cite{uspstf}. While several screening tools have been developed, such as Barium-meal gastric photofluorography~\cite{hamashima2008japanese}, upper endoscopy~\cite{jun2017effectiveness,choi2012performance,hu2021identifying}, and serum pepsinogen levels~\cite{miki2006gastric}, they are challenging to apply to the general population due to their invasiveness, moderate sensitivity/specificity, high cost, or side effects. Therefore, there is an urgent need for novel screening methods that are noninvasive, highly accurate, low-cost, and ready to distribute.

Non-contrast CT is a commonly used imaging protocol for various clinical purposes. It is a non-invasive, relatively low-cost, and safe procedure that exposes patients to less radiation dose and does not require the use of contrast injection that may cause serious side effects (compared to multi-phase contrast-enhanced CT). With recent advances in AI, opportunistic screening of diseases using non-contrast CT during routine clinical care performed for other clinical indications, such as lung and colorectal cancer screening, presents an attractive approach to early detect treatable and preventable diseases~\cite{pickhardt2022value}. However, whether early detection of gastric cancer using non-contrast CT scans is possible remains unknown. This is because early-stage gastric tumors may only invade the mucosal and muscularis layers, which are difficult to identify without the help of stomach preparation and contrast injection. Additionally, the poor contrast between the tumor and normal stomach wall/tissues on non-contrast CT scans and various shape alterations of gastric cancer, further exacerbates this challenge.  

In this paper, we propose a novel approach for detecting gastric cancer on non-contrast CT scans. Unlike the conventional ``segmentation for classification" methods that directly employ segmentation networks, we developed a cluster-induced Mask Transformer that performs segmentation and global classification simultaneously. Given the high variability in shape and texture of gastric cancer, we encode these features into learnable clusters and utilize cluster analysis during inference. By incorporating self-attention layers for global context modeling, our model can leverage both local and global cues for accurate detection. In our experiments, the proposed approach outperforms nnUNet~\cite{isensee2021nnu} by 0.032 in AUC, 5.0\% in sensitivity, and 4.1\% in specificity. These results demonstrate the potential of our approach for opportunistic screening of gastric cancer in asymptomatic patients using non-contrast CT scans.

\section{Related Work}
\noindent\textbf{Automated Cancer Detection.} Researchers have explored automated tumor detection techniques on endoscopic~\cite{luo2019real,li2020convolutional}, pathological images~\cite{song2020clinically}, and the prediction of cancer prognosis~\cite{li2022ct}. Recent developments in deep learning have significantly improved the segmentation of gastric tumors~\cite{li20213d}, which is critical for their detection. However, our framework is specifically designed for non-contrast CT scans, which is beneficial for asymptomatic patients. While previous studies have successfully detected pancreatic~\cite{xia2021effective} and esophageal~\cite{yao2022effective} cancers on non-contrast CT, identifying gastric cancer presents a unique challenge due to its subtle texture changes, various shape alterations, and complex background, \textit{e.g.}, irregular gastric wall; liquid and contents in the stomach. 

\noindent\textbf{Mask Transformers.} Recent studies have used Transformers for natural and medical image segmentation~\cite{tang2022self_swinunetr}. Mask Transformers~\cite{wang2021max,cheng2021per,cheng2022masked,Yuan_2023_CVPR} further enhance CNN-based backbones by incorporating stand-alone Transformer blocks, treating object queries in DETR~\cite{carion2020end} as memory-encoded queries for segmentation. CMT-Deeplab~\cite{yu2022cmt} and KMaX-Deeplab~\cite{yu2022k} have recently proposed interpreting the queries as clustering centers and adding regulatory constraints for learning the cluster representations of the queries. Mask Transformers are locally sensitive to image textures for precise segmentation and globally aware of organ-tumor morphology for recognition. Their cluster representations demonstrate a remarkable balance of intra-cluster similarity and inter-class discrepancy. Therefore, Mask Transformers are an ideal choice for an end-to-end joint segmentation and classification system for detecting gastric cancer.
\section{Methods}
\noindent\textbf{Problem Formulation.} Given a non-contrast CT scan, cancer screening is a binary classification with two classes as $\mathcal{L}=\{0, 1\}$, where 0 stands for``normal'' and 1 for``GC'' (gastric cancer). The entire dataset is denoted by $\mathcal{S} = \{(\mathbf{X}_i, \mathbf{Y}_i, \mathbf{P}_i) | i=1,2,\cdots,N\}$, where $\mathbf{X}_i$ is the $i$-th non-contrast CT volume, with $\mathbf{Y}_i$ being the voxel-wise label map of the same size as $\mathbf{X}_i$ and $K$ channels. Here, $K=3$ represents the background, stomach, and GC tumor. $\mathbf{P}_i \in \mathcal{L}$ is the class label of the image, confirmed by pathology, radiology, or clinical records. In the testing phase, only $\mathbf{X}_i$ is given, and our goal is to predict a class label for $\mathbf{X}_i$.  

\begin{figure}[ht]
    \centering
    \includegraphics[width=\linewidth]{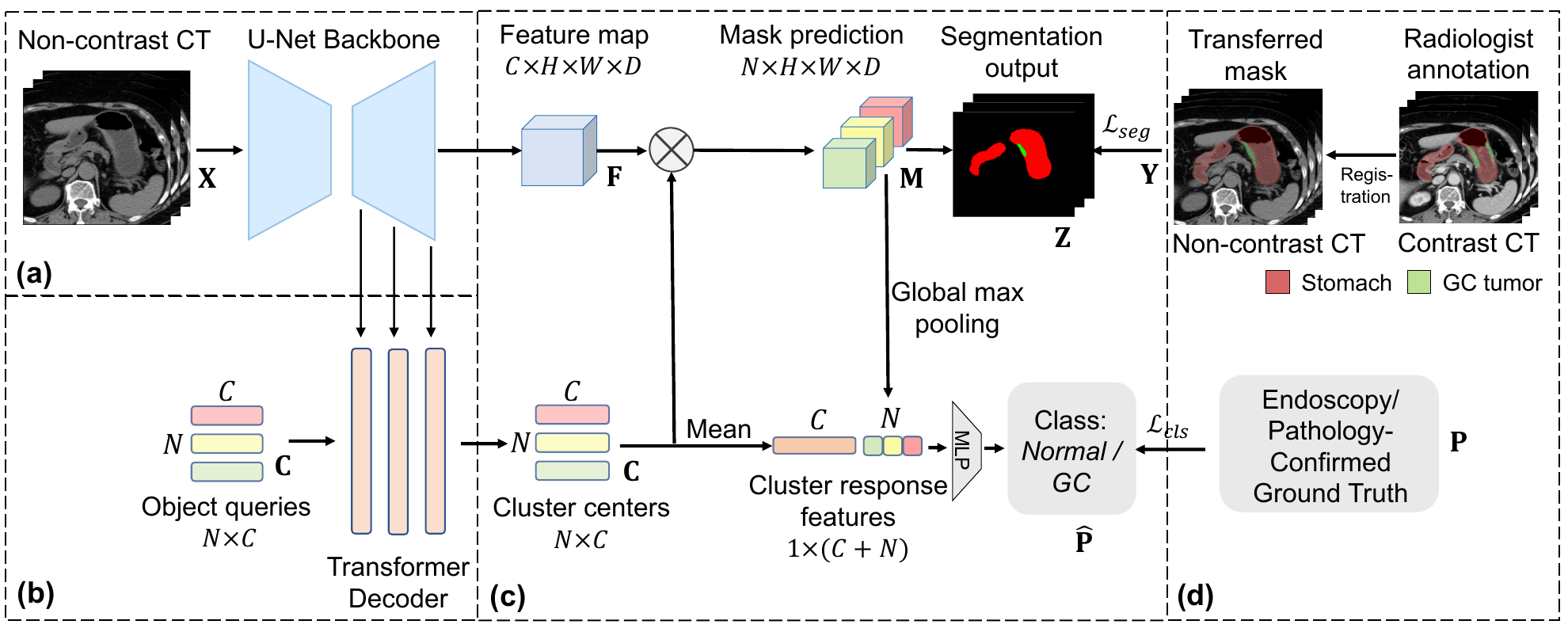}
    \caption{Method overview. (a) The non-contrast CT image is first forwarded into a U-Net~\cite{ronneberger2015u_unet0,isensee2021nnu} to extract a feature map. (b) Learnable object queries interact with the multi-level U-Net features through a Transformer Decoder and produce learned cluster centers. (c) All the pixels are assigned to cluster centers by matrix multiplication. The cluster assignment (\textit{a.k.a.} mask prediction) is further used to generate the final segmentation output and the classification probability. (d) The entire network is supervised by transferred masks from radiologists' annotation on contrast-enhanced CT and endoscopy or pathology-confirmed ground truth. }
    \vspace{-2mm}
    \label{fig:method}
\end{figure}

\noindent\textbf{Knowledge Transfer from Contrast-Enhanced to Non-contrast CT.} To address difficulties with tumor annotation on non-contrast CTs, the radiologists start by annotating a voxel-wise tumor mask on the contrast-enhanced CT, referring to clinical and endoscopy reports as needed. DEEDs~\cite{heinrich2013mrf} registration is then performed to align the contrast-enhanced CT with the non-contrast CT and the resulting deformation field is applied to the annotated mask. Any misaligned ones are revised manually. In this manner  (Fig.~\ref{fig:method}d), a relatively coarse yet highly reliable tumor mask can be obtained for the non-contrast CT image.

\noindent\textbf{Cluster-Induced Classification with Mask Transformers.}
Segmentation for classification is widely used in tumor detection~\cite{xia2021effective,yao2022effective,zhu2019multi}. We first train a UNet~\cite{ronneberger2015u_unet0,isensee2021nnu} to segment the stomach and tumor regions using the masks from the previous step. This UNet considers local information and can only extract stomach ROIs well during testing. However, local textures are inadequate for accurate gastric tumor detection on non-contrast CTs, so we need a network of both local sensitivity to textures and global awareness of the organ-tumor morphology. Mask transformer~\cite{wang2021max,cheng2021per,cheng2022masked} is a well-suited approach to boost the CNN backbone with stand-alone transformer blocks. Recent studies~\cite{yu2022cmt,yu2022k} suggest interpreting object queries as cluster centers, which naturally exhibit intra-cluster similarity and inter-class discrepancy. Inspired by this, we further develop a deep classification model on top of learnable cluster representations.

Specifically, given image $\mathbf{X} \in \mathbb{R}^{H \times W \times D}$, annotation $\mathbf{Y} \in \mathbb{R}^{K \times HWD}$, and patient class $\mathbf{P} \in \mathcal{L}$, our model consists of three components: 1) a CNN backbone to extract its pixel-wise features $\mathbf{F} \in \mathbb{R}^{C \times HWD}$ (Fig.~\ref{fig:method}a), 2) a transformer module (Fig.~\ref{fig:method}b), and 3) a multi-task cluster inference module(Fig.~\ref{fig:method}c). The transformer module gradually updates a set of randomly initialized object queries $\mathbf{C} \in \mathbb{R}^{N \times C}$, \textit{i.e.}, to meaningful mask embedding vectors through cross-attention between object queries and multi-scale pixel features,
\begin{equation}
    \mathbf{C} \leftarrow \mathbf{C} + \arg\max_{N} (\mathbf{Q}^c (\mathbf{K}^p)^{\mathrm{T}}) \mathbf{V}^p,
\end{equation}
where $c$ and $p$ stand for query and pixel features, $\mathbf{Q}^c, \mathbf{K}^p, \mathbf{V}^p$ represent linearly projected query, key, and value. We adopt cluster-wise argmax from KMax-DeepLab~\cite{yu2022k} to substitute spatial-wise softmax in the original settings.

We further interpret the object queries as cluster centers from a cluster analysis perspective. All the pixels in the convolutional feature map are assigned to different clusters based on these centers. The assignment of clusters (\textit{a.k.a.} mask prediction) $\mathbf{M} \in \mathbb{R}^{N \times HWD}$ is computed as the cluster-wise softmax function over the matrix product between the cluster centers $\mathbf{C}$ and pixel-wise feature matrix $\mathbf{F}$, \textit{i.e.}, 
\begin{equation}
    \mathbf{M} = \operatorname{Softmax}_{N}(\mathbf{R}) = \operatorname{Softmax}_{N}(\mathbf{C}\mathbf{F}).
\end{equation}
The final segmentation logits $\mathbf{Z} \in \mathbb{R}^{K \times HWD}$ are obtained by aggregating the pixels within each cluster according to cluster-wise classification, which treats pixels within a cluster as a whole. The aggregation of pixels is achieved by $\mathbf{Z} = \mathbf{C}_K \mathbf{M}$, where the cluster-wise classification $\mathbf{C}_K$ is represented by an MLP that projects the cluster centers $\mathbf{C}$ to $K$ channels (the number of segmentation classes).

The learned cluster centers possess high-level semantics with both inter-cluster discrepancy and intra-cluster similarity for effective classification. Rather than directly classifying the final feature map, we first generate the cluster-path feature vector by taking the channel-wise average of cluster centers $\Bar{\mathbf{C}} = \frac{1}{N}\sum_{i=1} \mathbf{C}_i \in \mathbb{R}^{C}$. Additionally, to enhance the consistency between the segmentation and classification outputs, we apply global max pooling to cluster assignments $\mathbf{R}$ to obtain the pixel-path feature vector $\Bar{\mathbf{R}} \in \mathbb{R}^{N}$. This establishes a direct connection between classification features and segmentation predictions. Finally, we concatenate these two feature vectors to obtain the final feature and project it onto the classification prediction $\Hat{\mathbf{P}} \in \mathbb{R}^2$ via a two-layer MLP.

The overall training objective is formulated as,
\begin{equation}
    \mathcal{L} = \mathcal{L}_{seg}(\mathbf{Z}, \mathbf{Y}) + \mathcal{L}_{cls}(\Hat{\mathbf{P}}, \mathbf{P}),
\end{equation}
where the segmentation loss $\mathcal{L}_{seg}(\cdot,\cdot)$ is a combination of Dice and cross entropy losses, and the classification loss $\mathcal{L}_{cls}(\cdot,\cdot)$ is cross entropy loss. 

\section{Experiments}
\subsection{Experimental setup}
\noindent\textbf{Dataset and Ground Truth.} Our study analyzed a dataset of CT scans collected from Guangdong Province People's Hospital between years 2018 and 2020, with 2,139 patients consisting of 787 gastric cancer and 1,352 normal cases. We used the latest patients in the second half of 2020 as a hold-out test set, resulting in a training set of 687 gastric cancer and 1,204 normal cases, and a test set of 100 gastric cancer and 148 normal cases. We randomly selected 20\% of the training data as an internal validation set. To further evaluate specificity in a larger population, we collected an external test set of 903 normal cases from Shengjing Hospital. Cancer cases were confirmed through endoscopy (and pathology) reports, while normal cases were confirmed by radiology reports and a two-year follow-up. All patients underwent multi-phase CTs with a median spacing of $0.75 \times 0.75 \times 5.0$ mm and an average size of (512, 512, 108) voxel. Tumors were annotated on the venous phase by an experienced radiologist specializing in gastric imaging using CTLabeler~\cite{wang2023cascaded}, while the stomach was automatically annotated using a self-learning model~\cite{zhang2018self}.

\begin{figure}[t]
\centering  
\begin{subfigure}[b]{0.55\linewidth}
     \centering
     \includegraphics[width=\linewidth]{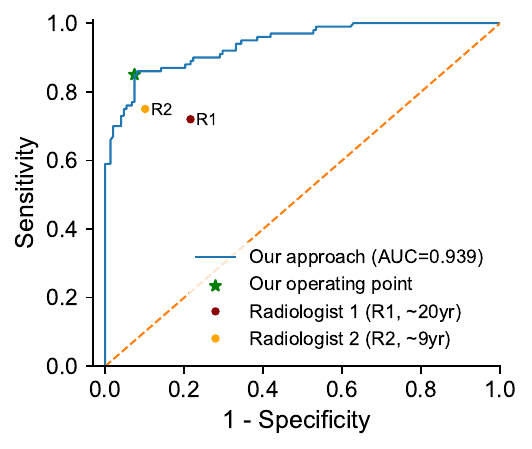}
     \caption{ROC curve}
     \label{fig:roc}
 \end{subfigure}     
\begin{subfigure}[b]{0.4\linewidth}
         \centering
         \includegraphics[width=\linewidth]{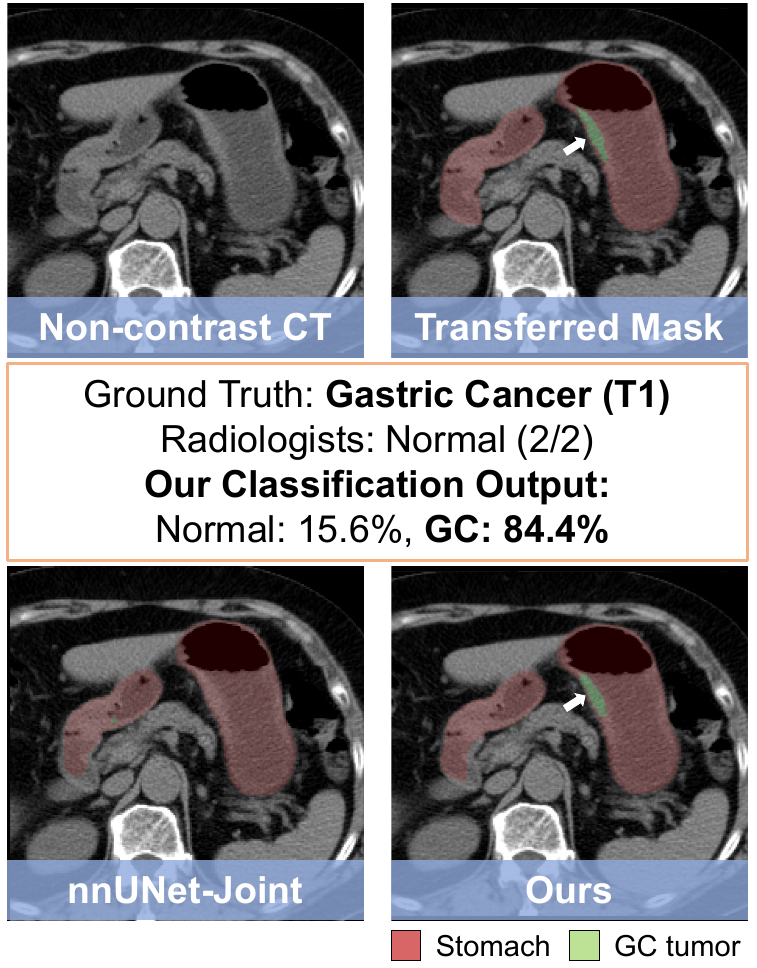}
         \caption{A visual example}
         \label{fig:case_study}
     \end{subfigure}
\caption{(a) ROC curve for our model versus two experts on the hold-out test set of $n=248$ patients for binary classification. (b) A visual example in the test set. This early-stage GC case is miss-detected by both radiologists and nnUNet~\cite{isensee2021nnu} but our model succeeds to locate the tumor.}
  \label{fig:auc_case} 
  \vspace{-5mm}
\end{figure}

\noindent\textbf{Implementation Details.} We resampled each CT volume to the median spacing while normalizing it to have zero mean and unit variance. During training, we cropped the 3D bounding box of the stomach and added a small margin of (32, 32, 4). We used nnUNet~\cite{isensee2021nnu} as the backbone, with four transformer decoders, each taking pixel features with output strides of 32, 16, 8, and 4. We set the number of object queries $N$ to 8, with each having a dimension of 128, and included an eight-head self-attention layer in each block. The patch size used during training and inference is (192, 224, 40) voxel. We followed~\cite{isensee2021nnu} to augment data. We trained the model with RAdam using a learning rate of $10^{-4}$ and a (backbone) learning rate multiplier of 0.1 for 1000 epochs, with a frozen backbone of the pre-trained nnUNet~\cite{isensee2021nnu} for the first 50 epochs. To enhance performance, we added deep supervision by aligning the cross-attention map with the final segmentation map, as per KMax-Deeplab~\cite{yu2022cmt}. The hidden layer dimension in the two-layer MLP is 128. We also trained a standard UNet~\cite{ronneberger2015u_unet0,isensee2021nnu} to localize the stomach region in the entire image in the testing phase.

\noindent\textbf{Evaluation Metrics and Reader Study.} For the binary classification, model performance is evaluated using area under ROC curve (AUC), sensitivity (Sens.), and specificity (Spec.). And successful localization of the tumors is considered when the overlap between the segmentation mask generated by the model and the ground truth is greater than 0.01, measured by the Dice score. A reader study was conducted with two experienced radiologists, one from Guangdong Province People's Hospital with 20 years of experience and the other from The First Affiliated Hospital of Zhejiang University with 9 years of experience in gastric imaging. The readers were given 248 non-contrast CT scans from the test set and asked to provide a binary decision for each scan, indicating whether the scan showed gastric cancer. No patient information or records were provided to the readers. Readers were informed that the dataset might contain more tumor cases than the standard prevalence observed in screening, but the proportion of case types was not disclosed. Readers used ITK-SNAP~\cite{yushkevich2006user} to interpret the CT scans without any time constraints.
\begin{table*}[t]
\centering
\setlength{\tabcolsep}{3pt}
\caption{Results on binary classification: gastric cancer vs. normal. The 95\% confidence interval of each metric is listed. $\dagger$: $p < 0.05$ for DeLong test (ours vs. nnUNet-S4C). *: $p < 0.05$ for permutation test (ours vs. nnUNet-S4C and radiologist experts). Sens.: Sensitivity. Spec.: Specificity.}
\begin{tabular}{l|c|c|c|c}
\hline
&\multicolumn{3}{c|}{Internal Hold-out ($n=248$)} & External ($n=903$) \\
\hline
Method            & AUC & Sens.(\%) & Spec.(\%) & Spec.(\%) \\ \hline
Mean of radiologists &   -  &  73.5     &  84.1     &    -        \\ \hline
nnUNet-S4C~\cite{isensee2021nnu}        & 0.907    & 80.0     &  88.5     &      96.6      \\
                 & (0.862, 0.942)     &  (72.0, 87.5)     & (83.3, 93.5)      &    (95.2, 97.8)        \\ \hline
TransUNet-S4C~\cite{chen2021transunet}     &  0.916   &  82.0     & 90.5      &     96.0       \\
                  & (0.876, 0.952)    & (74.7, 89.5)      &  (86.1, 94.8)     &   (94.8, 97.2)         \\ \hline
nnUNet-Joint~\cite{isensee2021nnu}     &  0.924   & 81.0      & 90.5      &    97.6        \\
                  & (0.885, 0.959)    &  (73.0, 87.9)     &  (85.1, 95.0)     &      (96.5, 98.6)      \\ \hline
\textbf{Ours}              & $\textbf{0.939}^{\dagger}$    &  $\textbf{85.0}^{*}$    & $\textbf{92.6}^{*}$      & \textbf{97.7}           \\
             &  (0.910, 0.964)     &  (78.1, 91.1)     & (88.0, 96.5)      &   (96.7, 98.7)         \\ \hline
\end{tabular}
\label{tab:auc}
\vspace{-6mm}
\end{table*}

\noindent\textbf{Compared Baselines.} \autoref{tab:auc} presents a comparative analysis of our proposed method with three baselines. The first two approaches belong to ``Segmentation for classification" (S4C)~\cite{zhu2019multi,yao2022effective}, using nnUNet~\cite{isensee2021nnu} and TransUNet~\cite{chen2021transunet}. A case is classified as positive if the segmented tumor volume exceeds a threshold that maximizes the sum of sensitivity and specificity on the validation set. The third baseline (denoted as ``nnUNet-Joint") integrates a CNN classification head into UNet~\cite{isensee2021nnu} and trained end-to-end. We obtain the 95\% confidence interval of AUC, sensitivity, and specificity values from 1000 bootstrap replicas of the test dataset for statistical analysis. For statistical significance, we conduct a DeLong test between two AUCs (ours vs. compared method) and a permutation test between two sensitivities or specificities (ours vs. compared method and radiologists). 

\subsection{Results}
\begin{table*} [t]
\caption{Patient-level detection and tumor-level localization results (\%) over gastric cancer across different T-stages. Tumor-level localization evaluates how segmented masks overlap with the ground-truth cancer (Dice $>$ 0.01 for correct detection). Miss-T: Missing of T stage information.}
\centering
\begin{tabular}{l| c  |   c|c|c|c|c }
  \hline
Method & Criteria     & T1 &T2 & T3& T4 & Miss-T\\
\hline
nnUNet-Joint~\cite{isensee2021nnu} & Patient  & 30.0(3/10) &66.7(6/9)  &94.1(32/34)  &100.0(9/9) & 86.1(31/36)\\
&  Tumor  & 20.0(2/10) &55.6(5/9)  & 94.1(32/34) &100.0(9/9)  &80.6(29/36) \\
 \hline
  \textbf{Ours} & Patient  & 60.0(6/10) & 77.8(7/9) & 94.1(32/34) &100.0(9/9) &86.1(31/36) \\
  & Tumor  & 30.0(3/10) &66.7(6/9) & 94.1(32/34) &100.0(9/9) & 80.6(30/36) \\
  \hline
 Radiologist 1  & Patient  & 50.0(5/10)&55.6(5/9)  &76.5(26/34)  & 88.9(8/9) &77.8(28/36) \\
\hline
 Radiologist 2  & Patient  & 30.0(3/10)&55.6(5/9)  &85.3(29/34)  & 100.0(9/9) &80.6(29/36) \\
\hline
\end{tabular}
\label{tab:det}
\vspace{-7mm}
\end{table*}

\begin{table*} [t]
\caption{Comparison with a state-of-the-art blood test on gastric cancer detection~\cite{klein2021clinical}, UGIS and endoscopy screening performance in large population~\cite{choi2012performance}, and early stage gastric cancer detection rate of senior radiologists on narrow-band imaging with magnifying endoscopy (ME-NBI)~\cite{hu2021identifying}. ($*$: We leave out two tumors \textit{in situ} within the test set in accordance with the setting in~\cite{klein2021clinical}.  $\dagger$: We also merely consider early-stage gastric cancer cases, including Tumor \textit{in situ}, T1, and T2 stages, among whom we successfully detect 17 of 19 cases.)}
\centering
\begin{tabular}{l|cc|c}
\hline
Method & Spec.(\%) & Sens.(\%) & \thead{Our sensitivity(\%) \\at the same specificity}           \\ \hline
Blood Test~\cite{klein2021clinical} & 99.5 & 66.7 &     $69.4^{*}$             \\ \hline
Upper-gastrointestinal series~\cite{choi2012performance} & 96.1 & 36.7 &     85.0            \\ \hline
Endoscopy screening~\cite{choi2012performance} & 96.0 & 69.0 &       85.0         \\ \hline
ME-NBI (early-stage) ~\cite{hu2021identifying} & 74.2 & 76.7 & $89.5^{\dagger}$  \\ \hline
\end{tabular}
\label{tab:sota}
\vspace{-5mm}
\end{table*}
\noindent\textbf{Our method Outperforms Baselines.} Our method outperforms three baselines (\autoref{tab:auc}) in all metrics, particularly in AUC and sensitivity. The advantage of our approach is that it captures the local and global information simultaneously in virtue of the unique architecture of mask transformer. It also extracts high-level semantics from cluster representations, making it suitable for classification and facilitating a holistic decision-making process. Moreover, our method reaches a considerable specificity of 97.7\% on the external test set, which is crucial in opportunistic screening for less false positives and unnecessary human workload.

\noindent\textbf{AI Models Surpass Experienced Radiologists on Non-contrast CT Scans.} As shown in \autoref{fig:roc}, our AI model's ROC curve is superior to that of two experienced radiologists. The model achieves a sensitivity of 85.0\% in detecting gastric cancer, which significantly exceeds the mean performance of doctors (73.5\%) and also surpasses the best performing doctor (R2: 75.0\%), while maintaining a high specificity. A visual example is presented in \autoref{fig:case_study}. This early-stage cancer (T1) is miss-detected by both radiologists, whereas classified and localized precisely by our model.

\noindent\textbf{Subgroup Analysis.} In \autoref{tab:det}, we report the performance of patient-level detection and tumor-level localization stratified by tumor (T) stage. We compare our model's performance with that of both radiologists. The results show that our model performs better in detecting early stage tumors (T1, T2) and provides more precise tumor localization. Specifically, our model detects 60.0\% (6/10) T1 cancers, and 77.8\% (7/9) T2 cancers, surpassing the best performing expert (50\% T1, 55.6\% T2). Meanwhile, our model maintains a reliable detection rate and credible localization accuracy for T3 and T4 tumors (2 of 34 T3 tumors missed).

\noindent\textbf{Comparison with Established Screening Tools.} Our method surpasses or performs on par with established screening tools~\cite{klein2021clinical,choi2012performance,hu2021identifying} in terms of sensitivity for gastric cancer detection at a similar specificity level with a relatively large testing patient size ($n=1151$ by integrating the internal and external test sets), as shown in \autoref{tab:sota}. This finding sheds light on the opportunity to employ automated AI systems to screen gastric cancer using non-contrast CT scans. 

\section{Conclusion}
We propose a novel Cluster-induced Mask Transformer for gastric cancer detection on non-contrast CT scans. Our approach outperforms strong baselines and experienced radiologists. Compared to other screening methods, such as blood tests, endoscopy, upper-gastrointestinal series, and ME-NBI, our approach is non-invasive,  cost-effective, safe, and more accurate for detecting early-stage tumors. The robust performance of our approach demonstrates its potential for opportunistic screening of gastric cancer in the general population.

\vspace{0.2cm}
\noindent\textbf{Acknowledgement}
This work was supported by Alibaba Group through Alibaba Research Intern Program. Bin Dong and Li Zhang was partly supported by NSFC 12090022 and 11831002, and Clinical Medicine Plus X-Young Scholars Project of Peking University PKU2023LCXQ041.
\bibliography{paper680}

\begin{thebibliography}{10}
\providecommand{\url}[1]{\texttt{#1}}
\providecommand{\urlprefix}{URL }
\providecommand{\doi}[1]{https://doi.org/#1}

\bibitem{carion2020end}
Carion, N., Massa, F., Synnaeve, G., Usunier, N., Kirillov, A., Zagoruyko, S.:
  End-to-end object detection with transformers. In: ECCV. pp. 213--229.
  Springer (2020)

\bibitem{chen2021transunet}
Chen, J., Lu, Y., Yu, Q., Luo, X., Adeli, E., Wang, Y., Lu, L., Yuille, A.L.,
  Zhou, Y.: Transunet: Transformers make strong encoders for medical image
  segmentation. arXiv preprint arXiv:2102.04306  (2021)

\bibitem{cheng2022masked}
Cheng, B., Misra, I., Schwing, A.G., Kirillov, A., Girdhar, R.:
  Masked-attention mask transformer for universal image segmentation. In: CVPR.
  pp. 1290--1299 (2022)

\bibitem{cheng2021per}
Cheng, B., Schwing, A., Kirillov, A.: Per-pixel classification is not all you
  need for semantic segmentation. In: NeurIPS. vol.~34, pp. 17864--17875 (2021)

\bibitem{choi2012performance}
Choi, K.S., Jun, J.K., Park, E.C., Park, S., Jung, K.W., Han, M.A., Choi, I.J.,
  Lee, H.Y.: Performance of different gastric cancer screening methods in
  korea: a population-based study. PLoS One  \textbf{7}(11),  e50041 (2012)

\bibitem{hamashima2008japanese}
Hamashima, C., Shibuya, D., Yamazaki, H., Inoue, K., Fukao, A., Saito, H.,
  Sobue, T.: The japanese guidelines for gastric cancer screening. Japanese
  journal of clinical oncology  \textbf{38}(4),  259--267 (2008)

\bibitem{heinrich2013mrf}
Heinrich, M.P., Jenkinson, M., Brady, M., Schnabel, J.A.: Mrf-based deformable
  registration and ventilation estimation of lung ct. IEEE transactions on
  medical imaging  \textbf{32}(7),  1239--1248 (2013)

\bibitem{hu2021identifying}
Hu, H., Gong, L., Dong, D., Zhu, L., Wang, M., He, J., Shu, L., Cai, Y., Cai,
  S., Su, W., et~al.: Identifying early gastric cancer under magnifying
  narrow-band images with deep learning: a multicenter study. Gastrointestinal
  Endoscopy  \textbf{93}(6),  1333--1341 (2021)

\bibitem{isensee2021nnu}
Isensee, F., Jaeger, P.F., Kohl, S.A., Petersen, J., Maier-Hein, K.H.: nnu-net:
  a self-configuring method for deep learning-based biomedical image
  segmentation. Nature methods  \textbf{18}(2),  203--211 (2021)

\bibitem{jun2017effectiveness}
Jun, J.K., Choi, K.S., Lee, H.Y., Suh, M., Park, B., Song, S.H., Jung, K.W.,
  Lee, C.W., Choi, I.J., Park, E.C., et~al.: Effectiveness of the korean
  national cancer screening program in reducing gastric cancer mortality.
  Gastroenterology  \textbf{152}(6),  1319--1328 (2017)

\bibitem{klein2021clinical}
Klein, E., Richards, D., Cohn, A., Tummala, M., Lapham, R., Cosgrove, D.,
  Chung, G., Clement, J., Gao, J., Hunkapiller, N., et~al.: Clinical validation
  of a targeted methylation-based multi-cancer early detection test using an
  independent validation set. Annals of Oncology  \textbf{32}(9),  1167--1177
  (2021)

\bibitem{li20213d}
Li, H., Liu, B., Zhang, Y., Fu, C., Han, X., Du, L., Gao, W., Chen, Y., Liu,
  X., Wang, Y., et~al.: 3d ifpn: Improved feature pyramid network for automatic
  segmentation of gastric tumor. Frontiers in oncology  \textbf{11},  618496
  (2021)

\bibitem{li2022ct}
Li, J., Chen, Z., Chen, Y., Zhao, J., He, M., Li, X., Zhang, L., Dong, B.,
  Zhang, X., Tang, L., et~al.: Ct-based delta radiomics in predicting the
  prognosis of stage iv gastric cancer to immune checkpoint inhibitors.
  Frontiers in Oncology  \textbf{12} (2022)

\bibitem{li2020convolutional}
Li, L., Chen, Y., Shen, Z., Zhang, X., Sang, J., Ding, Y., Yang, X., Li, J.,
  Chen, M., Jin, C., et~al.: Convolutional neural network for the diagnosis of
  early gastric cancer based on magnifying narrow band imaging. Gastric Cancer
  \textbf{23},  126--132 (2020)

\bibitem{luo2019real}
Luo, H., Xu, G., Li, C., He, L., Luo, L., Wang, Z., Jing, B., Deng, Y., Jin,
  Y., Li, Y., et~al.: Real-time artificial intelligence for detection of upper
  gastrointestinal cancer by endoscopy: a multicentre, case-control, diagnostic
  study. The Lancet Oncology  \textbf{20}(12),  1645--1654 (2019)

\bibitem{miki2006gastric}
Miki, K.: Gastric cancer screening using the serum pepsinogen test method.
  Gastric cancer  \textbf{9},  245--253 (2006)

\bibitem{seer2022cancer}
National Cancer~Institute, S.P.: Cancer stat facts: Stomach cancer.
  \url{https://seer.cancer.gov/statfacts/html/stomach.html} (2023)

\bibitem{pickhardt2022value}
Pickhardt, P.J.: Value-added opportunistic {CT} screening: state of the art.
  Radiology  \textbf{303}(2),  241--254 (2022)

\bibitem{ronneberger2015u_unet0}
Ronneberger, O., Fischer, P., Brox, T.: U-net: Convolutional networks for
  biomedical image segmentation. In: MICCAI. pp. 234--241. Springer (2015)

\bibitem{smyth2020gastric}
Smyth, E.C., Nilsson, M., Grabsch, H.I., van Grieken, N.C., Lordick, F.:
  Gastric cancer. The Lancet  \textbf{396}(10251),  635--648 (2020)

\bibitem{song2020clinically}
Song, Z., Zou, S., Zhou, W., Huang, Y., Shao, L., Yuan, J., Gou, X., Jin, W.,
  Wang, Z., Chen, X., et~al.: Clinically applicable histopathological diagnosis
  system for gastric cancer detection using deep learning. Nature
  communications  \textbf{11}(1), ~4294 (2020)

\bibitem{tang2022self_swinunetr}
Tang, Y., Yang, D., Li, W., Roth, H.R., Landman, B., Xu, D., Nath, V.,
  Hatamizadeh, A.: Self-supervised pre-training of swin transformers for 3d
  medical image analysis. In: CVPR. pp. 20730--20740 (2022)

\bibitem{uspstf}
USPSTF: {U.S. Preventive Services Task Force, Recommendations}.
  \url{https://www.uspreventiveservicestaskforce.org/uspstf/topic\_search\_results?topic\_status=P}
  (2023)

\bibitem{wang2023cascaded}
Wang, F., Cheng, C.T., Peng, C.W., Yan, K., Wu, M., Lu, L., Liao, C.H., Zhang,
  L.: A cascaded approach for ultraly high performance lesion detection and
  false positive removal in liver ct scans. arXiv preprint arXiv:2306.16036
  (2023)

\bibitem{wang2021max}
Wang, H., Zhu, Y., Adam, H., Yuille, A., Chen, L.C.: Max-deeplab: End-to-end
  panoptic segmentation with mask transformers. In: CVPR. pp. 5463--5474 (2021)

\bibitem{xia2021effective}
Xia, Y., Yao, J., Lu, L., Huang, L., Xie, G., Xiao, J., Yuille, A., Cao, K.,
  Zhang, L.: Effective pancreatic cancer screening on non-contrast ct scans via
  anatomy-aware transformers. In: MICCAI. pp. 259--269. Springer (2021)

\bibitem{yao2022effective}
Yao, J., Ye, X., Xia, Y., Zhou, J., Shi, Y., Yan, K., Wang, F., Lin, L., Yu,
  H., Hua, X.S., et~al.: Effective opportunistic esophageal cancer screening
  using noncontrast ct imaging. In: MICCAI. pp. 344--354. Springer (2022)

\bibitem{yu2022cmt}
Yu, Q., Wang, H., Kim, D., Qiao, S., Collins, M., Zhu, Y., Adam, H., Yuille,
  A., Chen, L.C.: Cmt-deeplab: Clustering mask transformers for panoptic
  segmentation. In: CVPR. pp. 2560--2570 (2022)

\bibitem{yu2022k}
Yu, Q., Wang, H., Qiao, S., Collins, M., Zhu, Y., Adam, H., Yuille, A., Chen,
  L.C.: k-means mask transformer. In: ECCV. pp. 288--307. Springer (2022)

\bibitem{Yuan_2023_CVPR}
Yuan, M., Xia, Y., Dong, H., Chen, Z., Yao, J., Qiu, M., Yan, K., Yin, X., Shi,
  Y., Chen, X., Liu, Z., Dong, B., Zhou, J., Lu, L., Zhang, L., Zhang, L.:
  Devil is in the queries: Advancing mask transformers for real-world medical
  image segmentation and out-of-distribution localization. In: CVPR. pp.
  23879--23889 (2023)

\bibitem{yushkevich2006user}
Yushkevich, P.A., Piven, J., Hazlett, H.C., Smith, R.G., Ho, S., Gee, J.C.,
  Gerig, G.: User-guided 3d active contour segmentation of anatomical
  structures: significantly improved efficiency and reliability. Neuroimage
  \textbf{31}(3),  1116--1128 (2006)

\bibitem{zhang2018self}
Zhang, L., Gopalakrishnan, V., Lu, L., Summers, R.M., Moss, J., Yao, J.:
  Self-learning to detect and segment cysts in lung ct images without manual
  annotation. In: ISBI. pp. 1100--1103 (2018)

\bibitem{zhu2019multi}
Zhu, Z., Xia, Y., Xie, L., Fishman, E.K., Yuille, A.L.: Multi-scale
  coarse-to-fine segmentation for screening pancreatic ductal adenocarcinoma.
  In: MICCAI. pp. 3--12. Springer (2019)

\end{thebibliography}
\bibliographystyle{splncs04}

\end{document}